\documentstyle[11pt]{article}
\def\tr{{\rm tr}\;}
\def\l{\lambda}
\def\bphi{\:\mbox{\large\boldmath $\phi$}\,}
\def\t{\:{\sf t}\,}
\def\w{\:{\sf w}\,}
\def\W{\:{\sf W}\,}
\def\u{\,{\sf u}\,}
\def\v{\,{\sf v}\,}
\def\c{{\sf c}}
\def\q{{\tt q}\,}
\def\p{\varsigma\,}
\def\ua{u_{_{\rm A}}}
\def\ub{u_{_{\rm B}}}
\def\uc{u_{_{\rm C}}}
\def\ud{u_{_{\rm D}}}
\def\fa{f_{_{\rm A}}}
\def\fb{f_{_{\rm B}}}
\def\fc{f_{_{\rm C}}}
\def\fd{f_{_{\rm D}}}
\def\phia{\phi_{_{\rm A}}}
\def\phib{\phi_{_{\rm B}}}
\def\phic{\phi_{_{\rm C}}}
\def\phid{\phi_{_{\rm D}}}
\def\vpa{\varphi_{_{\rm A}}}
\def\vpb{\varphi_{_{\rm B}}}
\def\vpc{\varphi_{_{\rm C}}}
\def\vpd{\varphi_{_{\rm D}}}
\def\r{\:{\sf r}\,}
\def\wh{\:\hat{{\sf w}}\,}
\def\x{\,{\sf x}\,}

\def\xh{\,\hat{{\sf x}}\,}
\def\yh{\,\hat{{\sf y}}\,}
\def\fh{\,\hat{{\sf f}}\,}
\def\gh{\,\hat{{\sf g}}\,}
\def\U{\:{\sf U}\,}
\def\R{\:{\sf R}\,}
\def\Rh{\:\hat{{\sf R}}\,}
\def\Q{\:{\sf Q}\,}
\def\aR{{\cal R}}
\def\aQ{{\cal Q}}

\def\C{\:{\sf C}\,}
\def\Wh{\hat{W}}
\def\half{\frac{1}{2}}
\def\lm{{\textstyle \frac{\l}{\mu}}}

\def\mN{\!\!\!\!\!\pmod{N}}
\def\bma{\left(\begin{array}{cc}}
\def\ema{\end{array}\right)}
\def\op{\begin{picture}(12,21)
  \put(0,0){${\displaystyle\prod}$}
  \put(6,16){\circle{10}}
  \put(10.8,18){\vector(0,1){0}}
\end{picture}\;}

\begin{document}
\bigskip
\begin{center}
{\Large\bf
Quantum lattice KdV equation}\\
\bigskip
{\large\bf
     A. Yu. Volkov\footnote
     {On leave of absence from
Saint Petersburg Branch of the Steklov Mathematical Institute,\\
Fontanka 27, Saint Petersburg 191011, Russia}}\\
{\em Department of Theoretical Physics\\Uppsala University\\
     Box 803, S-751 08 Uppsala, Sweden}
\end{center}
\bigskip
\begin{quote}\begin{quote}
{\bf Abstract.} A quantum theory is developed for a difference-difference
system which can serve as a toy-model of the quantum
Korteveg-de-Vries equation.
\end{quote}\end{quote}

\section*{Introduction}

This Letter presents an example of a completely
integrable `discrete-space-time
quantum model' whose Heisenberg equations of motion have the form
$$ \bphi(\tau,n)\bphi(\tau,n-1)
                        +\l\bphi(\tau,n-1)\bphi(\tau-1,n-1)\;\;\;\;\;  $$
$$\;\;\;\;\; =\l\bphi(\tau,n)\bphi(\tau-1,n)
                             +\bphi(\tau-1,n)\bphi(\tau-1,n-1).\eqno(1)$$
By `discrete...model' we mean
\begin{itemize}
   \item[(i)] an algebra `of observables' $\Phi$,
\end{itemize}
whose generators $\bphi_n$ are labeled by integer numbers $n$ which are
regarded as a (discrete) spatial variable; together~with
\begin{itemize}
   \item[(ii)] an automorphism $\aQ$,
\end{itemize}
whose sequential action
$$              \x(0)\equiv\x\in\Phi                                   $$
$$    \aQ:\;\;\ldots\mapsto\x(\tau-1)
             \mapsto\x(\tau)\mapsto\x(\tau+1)\mapsto\ldots             $$
is viewed as the (discrete) time evolution. Thus, we intend to produce a
pair $\Phi\&\aQ$ such that the evolution $\bphi_n(\tau)$ of generators,
in the natural notation
$$                    \bphi(\tau,n)\equiv\bphi_n(\tau)   ,             $$
obeys the system (1).

Complete integrability is understood as the existence of a
commutative subalgebra `of conservation laws' preserved under
time evolution and spanning, in a sense, half of the algebra of
observables: it is commonly believed that a
Hamiltonian system may be either `completely' nonintegrable
possessing only a few conservation laws due to its manifest symmetries,
or completely integrable enjoying a whole lot of conservation laws,
one per degree of freedom. The commutative
subalgebra which we encounter in this Letter definitely contains
a lot of conservation laws but the question of how many is left to be
answered elsewhere.

Actually, we deal here not with a single model but rather with a
family of them $\Phi\&\aQ(\l)$, each model being related to a certain
value of a complex parameter $\l$ in (1).
Moreover, all their evolution automorphisms $\aQ(\l)$ turn
out to be mutually commuting and sharing the common subalgebra of
conservation laws.

The order of presentation is as follows.
In Section 1 we introduce an algebra of observables which is
basically the same lattice $U(1)$ exchange algebra
which appeared already in [FV93, 94].
Naturally, the behaviour of that algebra depends on the value of a
constant $\q$ involved in the commutation relations.
For simplicity we shall assume that $\q$ is a root of unity.

After some preliminaries (Sections 2 and 3) we pick in Section 4
from the algebra of observables a set
of `Fateev-Zamolodchikov $R$-matrices' $\R_n(\l)$ which satisfy a
chain of Yang-Baxter equations
$$   \R_{n-1}(\l)\R_n(\l\mu)\R_{n-1}(\mu)
                            =\R_n(\mu)\R_{n-1}(\l\mu)\R_n(\l)          $$
amounting to mutual commutativity
$$                    \Q(\l)\Q(\mu)=\Q(\mu)\Q(\l)                      $$
of their properly defined `ordered products'. The
family $\Q(\l)$ provides
the demanded commuting (inner) evolution automorphisms
$$               \bphi(\tau,n)\equiv\Q^{-\tau}\bphi_n \Q^{\tau}        $$
and doubles as their common conservation laws.

In Section 6 we eventually establish that these evolutions do solve the
equations (1). Prior to that, in Section 5, we discover
one more face of the family $\Q(\l)$. As a function of $\l$ it
proves to satisfy the Baxter equation [Bax]
$$   \q^{N\ell^2}\Q(\l)\t(\l)
                =\alpha^N (\l)\Q(\q^{-1}\l)+\delta^N (\l)\Q(\q\l)      $$
which in turn makes Bethe ansatz equations to emerge
in a purely algebraic context.

While the quantum system (1) seems to be a recent invention (it has
been looked at, albeit from a somewhat different angle, in [FV92, 94]),
its classical counterpart has been around for quite a while.
It was introduced (in a somewhat different form) by Hirota back
in 1977 [H] as an integrable difference-difference approximation of the
sine-Gordon equation but eventually proved far more universal making
perfect sense as a lattice counterpart of numerous integrable equations
including that of Korteveg and de Vries. To conclude the Introduction
we shall list various
continuous limits and alternative forms of the (classical) system (1).
This should give some idea of where our model fits into the scheme
of things accepted in Soliton Theory. For more of that and
a comprehensive list of relevant references see [NC].

Pairs of integers $(\tau,n)$ may naturally be viewed as vertices
of a plane lattice. In the shorthand notation
$$
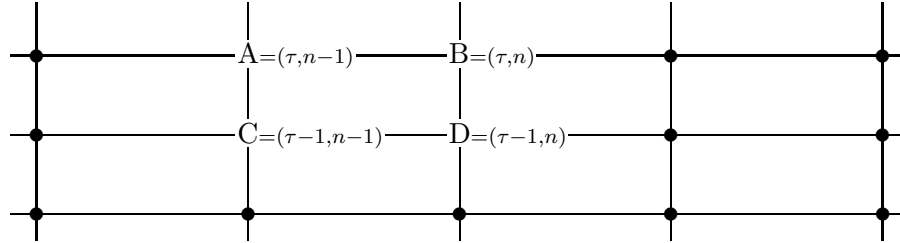
\begin{picture}(340,90)(-10,-10)
    \put(76,57){A$\scriptstyle =(\tau,n-1)$}
    \put(156,57){B$\scriptstyle =(\tau,n)$}
    \put(76,27){C$\scriptstyle =(\tau-1,n-1)$}
    \put(156,27){D$\scriptstyle =(\tau-1,n)$}
    \put(0,0){\circle*{5}}
    \put(80,0){\circle*{5}}
    \put(160,0){\circle*{5}}
    \put(240,0){\circle*{5}}
    \put(320,0){\circle*{5}}
    \put(0,30){\circle*{5}}
    \put(240,30){\circle*{5}}
    \put(320,30){\circle*{5}}
    \put(0,60){\circle*{5}}
    \put(240,60){\circle*{5}}
    \put(320,60){\circle*{5}}
    \put(-10,0){\line(1,0){340}}
    \put(-10,30){\line(1,0){85}}
    \put(155,30){\line(-1,0){23}}
    \put(330,30){\line(-1,0){129}}
    \put(-10,60){\line(1,0){85}}
    \put(155,60){\line(-1,0){34}}
    \put(330,60){\line(-1,0){141}}
    \put(0,-10){\line(0,1){90}}
    \put(80,-10){\line(0,1){35}}
    \put(80,55){\line(0,-1){19}}
    \put(80,80){\line(0,-1){14}}
    \put(160,-10){\line(0,1){35}}
    \put(160,55){\line(0,-1){19}}
    \put(160,80){\line(0,-1){14}}
    \put(240,-10){\line(0,1){90}}
    \put(320,-10){\line(0,1){90}}
                                                          \end{picture}$$
for a quartet of vertices enclosing some elementary cell of that lattice,
with subscripts instead of parenthesized arguments and without bold
letters reserved for the quantum case, the classical equations (1) read
$$ \phib\phia-\phid\phic-\l\left(\phib\phid-\phia\phic\right)=0.       $$

$\bullet$ At least one continuous limit is already quite apparent.
Let us put the lattice on the coordinate plane $(t,x)$ in such a way that
vertices $(\tau,n)$ go to points $(\l^{-1}\Delta \tau,\Delta n)$:
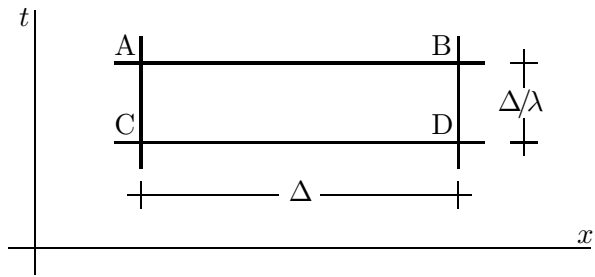
$$\begin{picture}(220,100)(-10,-10)
    \put(30,73){A}
    \put(150,73){B}
    \put(30,43){C}
    \put(150,43){D}
    \put(205,2){$x$}
    \put(-6,84){$t$}
    \put(96,18){$\Delta$}
    \put(175,52){$\Delta\!/\!\l$}
    \put(-10,0){\line(1,0){220}}
    \put(180,40){\line(1,0){10}}
    \put(180,70){\line(1,0){10}}
    \put(35,20){\line(1,0){57}}
    \put(108,20){\line(1,0){57}}
    \put(40,15){\line(0,1){10}}
    \put(160,15){\line(0,1){10}}
    \put(185,35){\line(0,1){14}}
    \put(185,61){\line(0,1){14}}
    \put(0,-10){\line(0,1){100}}
    \thicklines
    \put(30,40){\line(1,0){140}}
    \put(30,70){\line(1,0){140}}
    \put(40,30){\line(0,1){50}}
    \put(160,30){\line(0,1){50}}
                                                          \end{picture}$$
If now one manages to find a family of solutions to (1) depending on
the lattice spacing $\Delta$ and tending to a
smooth function $v(t,x)$ as $\Delta$ goes
to zero then that function can be easily seen to satisfy the equation
$$                      \phi_t-\phi_x=0.                               $$
This equation hardly needs any comment although at this point the
use of a nonlinear difference-difference equation to model a linear
differential one seems difficult to justify. Once, however, one takes
it as a Minkowsky version
of the Cauchy-Riemann equations things start to look like a unified
approach to conformal invariance and integrability.

$\bullet$ Let us now perform a more sophisticated
superimposition $(\tau,n)\longrightarrow
(\frac{\l-\l^3}{24}\:\Delta^3 \tau, \l\Delta\tau+\Delta n)$:
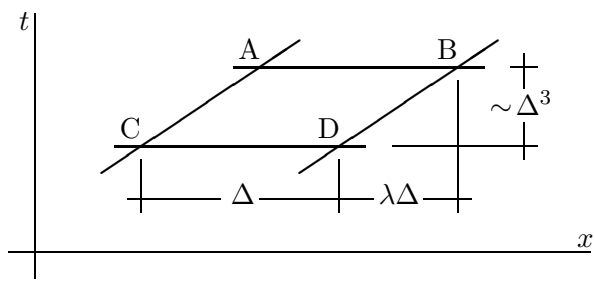
$$\begin{picture}(220,100)(-10,-10)
    \put(77,73){A}
    \put(152,73){B}
    \put(32,43){C}
    \put(107,43){D}
    \put(205,2){$x$}
    \put(-6,84){$t$}
    \put(74,18){$\Delta$}
    \put(130,18){$\l\Delta$}
    \put(172,52){$\sim\!\Delta^3$}
    \put(-10,0){\line(1,0){220}}
    \put(135,40){\line(1,0){55}}
    \put(180,70){\line(1,0){10}}
    \put(35,20){\line(1,0){35}}
    \put(85,20){\line(1,0){43}}
    \put(147,20){\line(1,0){18}}
    \put(40,15){\line(0,1){20}}
    \put(115,15){\line(0,1){20}}
    \put(160,15){\line(0,1){50}}
    \put(185,35){\line(0,1){14}}
    \put(185,62){\line(0,1){13}}
    \put(0,-10){\line(0,1){100}}
    \thicklines
    \put(30,40){\line(1,0){95}}
    \put(75,70){\line(1,0){95}}
    \put(25,30){\line(3,2){75}}
    \put(100,30){\line(3,2){75}}
                                                          \end{picture}$$
This time we come to the close relative of the KdV equation
$$    \phi_t+\phi_{xxx}-3\:\frac{\phi_{xx}\phi_x}{\phi}=0              $$
for any solution of which the potential
$$       u=\frac{\phi_{xx}}{\phi}                                      $$
solves the KdV equation itself
$$               u_t+u_{xxx}-6\: uu_x=0.                               $$

$\bullet$ To demonstrate a somewhat different scenario let us recall
how to turn (1) into the sine-Gordon equation. Before performing a
continuous limit we switch in (1) to the function $\varphi(\tau,n)$
$$       \phi(\tau,n)=e^{(-1)^n i \varphi(\tau,n)}                     $$
that brings (1) back to its original form
$$ \sin{\textstyle\half}(\vpa-\vpb-\vpc+\vpd)
                  +\l\sin{\textstyle\half}(\vpa+\vpb+\vpc+\vpd)=0  .   $$
Let the lattice cells now look like
$$
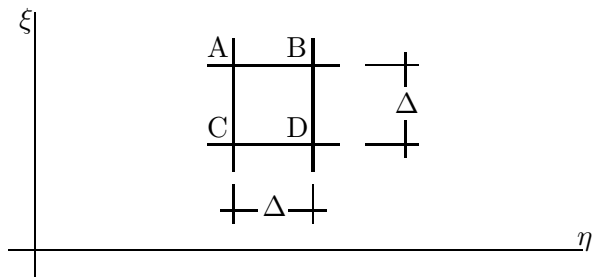
\begin{picture}(220,100)(-10,-10)
    \put(65,73){A}
    \put(95,73){B}
    \put(65,43){C}
    \put(95,43){D}
    \put(205,2){$\eta$}
    \put(-6,84){$\xi$}
    \put(86,13){$\Delta$}
    \put(136,52){$\Delta$}
    \put(-10,0){\line(1,0){217}}
    \put(125,40){\line(1,0){20}}
    \put(125,70){\line(1,0){20}}
    \put(70,15){\line(1,0){14}}
    \put(96,15){\line(1,0){14}}
    \put(75,10){\line(0,1){15}}
    \put(105,10){\line(0,1){15}}
    \put(140,35){\line(0,1){14}}
    \put(140,62){\line(0,1){13}}
    \put(0,-10){\line(0,1){100}}
    \thicklines
    \put(65,40){\line(1,0){50}}
    \put(65,70){\line(1,0){50}}
    \put(75,30){\line(0,1){50}}
    \put(105,30){\line(0,1){50}}
                                                          \end{picture}$$
and rescale
the constant in the equation making it dependent
on the lattice spacing
$$       \l=-\left( {\textstyle \frac{{\tt m}\Delta}{2}} \right)^2  .  $$
If now for a fixed value of the constant ${\tt m}$ one finds a family of
solutions $\varphi$ tending, as $\Delta\rightarrow 0$, to a smooth
function $\varphi(\xi,\eta)$ then this function satisfies the
sine-Gordon equation
$$  \varphi_{\xi\eta}
              +{\textstyle \frac{{\tt m}^2}{2}}\sin{2\varphi}=0   .    $$
Of course the original function $\phi$ can not survive under this
continuous limit becoming badly oscillating.

$\bullet$ Although the mere ability of (1) to unify KdV and SG equations
makes it a reasonable prospect, one is still left to wonder why
the would-be universal lattice equation does not look
special enough for that.
And indeed there exists a far more spectacular version of (1).
One can try to guess its form recollecting that the KdV equation
looks best in its Krichever-Novikov reincarnation\footnote{As this is
already the third form of the KdV equation we have met, so far, and the
arrival of a fourth one, the so-called modified KdV equation, is
imminent, let us recall how all these forms interact
$$\begin{array}{cll}
 &&\\
 f&&f_t+S[f]f_x=0 \\
 \downarrow&&\\
 \phi&=\frac{1}{\sqrt{f_x}}&\phi_t+\phi_{xxx}
                               -3\frac{\phi_{xx}\phi_x}{\phi}=0\\

 \downarrow&&\\
 p&=\frac{\phi_x}{\phi}
               =-\half\frac{f_{xx}}{f_x}&p_t+p_{xxx}-6p^2 p_x=0\\
 \downarrow&&\\
 u&=p^2+p_x=\frac{\phi_{xx}}{\phi}=-\half S[f]
                         \;\;\;\;\;\;\;\;\;&u_t+u_{xxx}-6uu_x=0\\
 &&

\end{array}$$
and for some while rename them from their traditional
names to $f/\psi/p/u$-equations correspondingly.}
$$                             f_t+S[f]f_x=0                           $$
where $S[f]$ stands for the Schwarz derivative
$$ S[f]=\frac{f_{xxx}}{f_x}
                    -\frac{3}{2}\left( \frac{f_{xx}}{f_x} \right)^2.   $$
Any decent lattice approximation for the $f$-equation should employ the
cross-ratio as a difference counterpart of the Schwarz derivative and
it is not difficult to spot the right one:
$$       \frac{(\fa-\fb)(\fc-\fd)}{(\fa-\fc)(\fb-\fd)}={\rm const}  .  $$
Indeed, this equation does produce, provided const$=\l^{-2}$, in the
above `parallelogram' continuous limit exactly the $f$-equation being as
well a completely integrable difference-difference model in its own
right. And just as the continuous $f$- and $\phi$-equations
are tied up by the Fuchs formula
$$                            f_x=\frac{1}{\phi^2}                     $$
their lattice counterparts are connected by the map defined by
$$         \frac{1}{\Delta}(f_n-f_{n-1})=\frac{1}{\phi_{n-1}\phi_n}    $$
where we omitted the argument $\tau$ common for all entries and moved the
remaining one to the subscript position.

The cross-ratio version will find extensive use in the forthcoming paper
addressing higher-order equations (KdV hierarchy), equations with two
fields (NLS hierarchy) and 2+1-dimensional equations (KP hierarchy).
Unfortunately, despite of its virtues this version has not yet been
really useful in the quantum theory where at present we are only able
to handle a free-field sort of algebra of observables associated
with the $\phi$-equation.

$\bullet$ By the way, at this point one finds himself well prepared
to design a lattice version of the $u$-equation. Introducing
$$              u_n=\frac{2\phi_n}{\phi_{n-1}+\phi_{n+1}}              $$
so that
$$      u_{n-1} u_n=4\:\frac{(f_{n+1}-f_n)(f_{n-1}-f_{n-2})}
                                    {(f_{n+1}-f_{n-1})(f_n-f_{n-2})}   $$
and
$$      u_{lat}\sim 1-{\textstyle \half}\Delta^2 u_{cont}              $$
one eventually transforms (1) into
$$  (\ua-\ud)(\ub-\uc)
               +{\textstyle\frac{1}{4}}(\l^{-2}-1)(\ua\ub-\uc\ud)^2=0  $$
approximating the KdV equation in its original form.

\section{Algebra of observables}

All the way through this Letter a constant $\q$ will
be an odd root of unity
$$                           \q^{2\ell+1}=1                            $$
and an integer number $N(\geq3)$ called the spatial period will be {\bf
odd}.
The only algebra of observables in use will be
the `exchange' algebra $\Phi$ with generators $\bphi_n$ subject to
\begin{itemize}
   \item[(i)] commutation relations
$$   \bphi_m\bphi_n=\q^{-\epsilon(m-n)}\bphi_n\bphi_m  ,               $$
with
$$    \epsilon(n)=1 \;\;\;\;n=1,3,5,\ldots,N-2                         $$
$$    \epsilon(n)=0 \;\;\;\;n=0,2,4,\ldots,N-1                         $$
$$    \epsilon(n+N)=\epsilon(n)+1,                                     $$
\end{itemize}
complemented by conditions
\begin{itemize}
   \item[(ii)]
$$                           \bphi_n^{2\ell+1}=1       ,               $$
   \item[(iii)]
$$    \bphi_n^{-1}\bphi_{n+N}^{} \;\;\mbox{does not depend on}\;\;n.   $$
\end{itemize}
The simplifying condition (ii) is natural if not really essential.
In (iii) one easily
recognizes a quasiperiodic boundary condition leaving in the algebra of
observables only $N$+1 independent generators\footnote
{$N$+1 is even, all right.}, for instance,~$\bphi_0,\bphi_1,
\ldots,\bphi_N$.

That algebra contains a useful `current' subalgebra $W$ with generators
$$             \w_n=\frac{\bphi_{n+1}}{\bphi_{n-1}}                    $$
that are easily seen to obey
\begin{itemize}
   \item[(i)] commutation relations
$$               \w_{n-1} \w_n=\q^2\w_n \w_{n-1}                       $$
$$               \w_m \w_n=\w_n \w_m    \;\;\; |m-n|\neq 1\mN   ,      $$
\end{itemize}
and conditions
\begin{itemize}
   \item[(ii)]
$$                      \w_n^{2\ell+1}=1  ,                            $$
   \item[(iii)]
$$                      \w_{n+N}=\w_n     .                            $$
\end{itemize}
Of course, (iii) is just the periodic boundary condition.
The whole algebra $\Phi$ is, in a sense, one degree of freedom larger
than that current subalgebra which has $N$ independent generators and
a central element
$$    \c=\q(\bphi_n^{-1}\bphi_{n+N}^{})^2
     =\q\w_1 \w_3 \ldots \w_N\w_2 \w_4 \ldots \w_{N-1}.                $$

A remark is in order here. The algebra of observables was designed
with the `second' periodic KdV bracket (aka the Virasoro algebra)
$$  {\textstyle\frac{1}{\gamma}}
     \{u(x),u(y)\}=  2(u(x)+u(y))\delta'(x-y)-\delta'''(x-y)           $$
in mind. Indeed, that bracket corresponds to
the sign-function-bracket (the exchange algebra)\footnote
{Strictly speaking, $\delta(\cdot)$ means here
the $2\pi$-periodic $\delta$-function while ${\rm sign}(\cdot)$ is
a function coinciding with the usual sign function
in the interval $[-\pi,\pi]$ and extending quasiperiodically
elsewhere: ${\rm sign}(x+2\pi)={\rm sign}(x)+2$.}
$$  {\textstyle\frac{1}{\gamma}}
 \{\phi(x),\phi(y)\}=-{\textstyle\half}\;{\rm sign}(x-y)\phi(x)\phi(y) $$
in the $\phi$-language and
the delta-prime-function-bracket (the current algebra)
$$  {\textstyle\frac{1}{\gamma}}
                    \{p(x),p(y)\}= \delta'(x-y)                        $$
in the $p$-language (see Introduction). Both are easily seen to be
classical ($\q=e^{i\hbar\gamma},\;
\hbar\rightarrow 0$) continuous ($\phi_{lat}\sim\phi_{cont},\;
w_{lat}\sim 1+2\Delta p_{cont}$) limits of the above commutation
relations. A natural question arises whether the Virasoro algebra itself
has a reasonable lattice counterpart. The answer seems to be
affirmative [FT, V, B, Fe] but this is another story.

\section{Fateev-Zamolodchikov $R$-matrix}

Let two operators $\u$ and $\v$ satisfy the condition
$$              \u^{2\ell+1}=\v^{2\ell+1}=1                            $$
and obey Weyl's commutation relation
$$                     \u\v = \q^2 \v\u.                               $$
Then, as was found in [FZ], the pair of functions of complex
variable taking values in the algebra generated by $\u$ and $\v$
$$    R_1(\l)=\:r(\l,\u)\;\;\;\;\;\;\;\;\;\;R_2(\l)=\:r(\l,\v),        $$
\begin{quote}
where
$$   r(\l,z)=\sum_{k=-\ell}^{\ell}\rho_k(\l)z^k                        $$
$$  \rho_k(\l)=\rho_{-k}(\l)=\q^{k^2}\prod_{j=1}^k(1-\l \q^{-2(j-1)})
\prod_{j=k+1}^{\ell}(1-\l \q^{2j})\;\;\;\;\;\;\;\;0\leq k \leq \ell    $$
and $\prod_1^0\ldots=\prod_{\ell+1}^{\ell}\ldots=1$,
\end{quote}
satisfies the `braid' Yang-Baxter equation
$$ R_1(\l)\:R_2(\l\mu)\:R_1(\mu)=R_2(\mu)\:R_1(\l\mu)\:R_2(\l)\:   .   $$

Before going on let us compile a list of some useful properties of the
function $r(\l,z)$. We shall often use them
in remaining sections, sometimes not mentioning it explicitly.
In what follows the second argument of $r$ is always assumed to
satisfy the condition
$$                   z^{2\ell+1}=1   .                                 $$
\begin{itemize}
   \item[(i)] $r(\l,z)$ is, up to a constant factor, the only
polynomial in $\l$
of degree $\ell$ satisfying the functional equation
$$     (\l+z)\:r(\l,\q z)=(1+\l z)\:r(\l,\q^{-1}z) .                   $$
   \item[(ii)] $r(\l,z)$ satisfies the functional equation
$$ (1+\l)(1+\l \q)\;r(\q\l,z)=(1+\l z)(1+\l z^{-1})\:r(\q^{-1}\l,z) .  $$
   \item[(iii)]
$$                       r(\l,z)=\:r(\l,z^{-1})                        $$
   \item[(iv)] the product of $r(\l,z)$ and $r(\l^{-1},z)$ does not
depend on $z$:
$$          r(\l,z)\:r(\l^{-1},z)=\varrho(\l)={\rm const}\cdot\l^{-\ell}
                  \prod_{{k=-\ell}\atop {k\neq 0}}^{\ell}(1+\l \q^k).  $$
    \item[(v)] At the point $\l=0$ the function $r(\l,z)$ turns into
a `truncated' theta-function
$$    r(0,z)=\theta(z)=\sum_{k=-\ell}^{\ell}\q^{k^2}z^k                $$
satisfying the functional equation
$$        z\:\theta(\q z)=\:\theta(\q^{-1}z)        .                  $$
The operators
$$  \Theta_1
     \equiv\:\theta(\u)\;\;\;\;\;\;\;\;\;\;\Theta_2\equiv\:\theta(\v)  $$
obey Artin's commutation relation
$$           \Theta_1\Theta_2\Theta_1=\Theta_2\Theta_1\Theta_2         $$
thus providing a `free-field' model of the braid group $B_2$.
    \item[(vi)]
$$                r(1,z)={\rm const}                                   $$
\end{itemize}
Some of these statements are quite transparent, some are less so. We will
present their proofs in a more detailed paper.

\section{Cyclic product}

Let $\Wh$ be a free algebra\footnote
{It is essential that the algebra $\Wh$, as opposed to the algebra $W$ of
Section 1, is devoid of the periodic boundary condition. It is not really
necessary to get rid of other relations defining $W$.}
with generators $\wh_n$.
Denote
\begin{itemize}
 \item[(i)] by $\p$ the homomorphism projecting $\Wh$ to $W$
$$                  \p(\wh_n)=\w_n                                     $$
$$                \p(\xh\yh)=\p(\xh)\:\p(\yh)  ,                       $$
 \item[(ii)] by $_{+N}$ the shift-by-period automorphism
of the algebra $\Wh$
$$           (\wh_n)_{+N}=\wh_{n+N}                                    $$
$$            (\xh\yh)_{+N}=\xh_{+N}\yh_{+N}           .               $$
\end{itemize}
Assign to monomials of $\Wh$ `overlap' numbers
$$  \nu\left(\wh_{n_1}^{k_1}\wh_{n_2}^{k_2}\ldots
             \wh_{n_L}^{k_L}\right)=\sum_{p=-\infty}^{+\infty}
      \sum_{i,j=1}^L p\: k_i k_j \:\delta_{\:n_j-n_i\:,\:pN-1}         $$
and utilize them in the coboundary which defines another
multiplication $\star$ in $\Wh$: for basis elements it reads
$$ \fh\star\gh=\q^{2\left(\nu(\fh\gh)-\nu(\fh)-\nu(\gh)\right)}\fh\gh  $$
and extends bilinearly elsewhere.
\bigskip
Then for any two elements $\xh,\yh$ of $\Wh$
$$                       \p(\xh\star\yh)=\p(\yh\star\xh_{+N}).         $$
We omit the proof for it is a straightforward computation.

The purpose of this $\p(\star)$ construction must be clear
for those familiar with the
Quantum Inverse Scattering Method: once we decide
to do without an `auxiliary space' and go for a purely algebraic version
of the $R$-matrix approach we need some direct method of computing
what used to be auxiliary space traces.

\section{Conservation laws}

We are going to prove that  `cyclic ordered products'
$$ \U(\l)\equiv\p\left( \Rh_1(\l)\star \Rh_2(\l)\star
              \ldots \star\Rh_N(\l)\right)                             $$
$$         =\sum_{j,k=-\ell}^{\ell} \q^{2jk}\;(\rho_j(\l)\w_{1}^j)\;
        \R_{2}(\l)\R_{3}(\l)\ldots \R_{N-1}(\l)\;(\rho_k(\l)\w_{N}^k)  $$
of FZ $R$-matrices
$$ \R_n(\l)
    \equiv\:r(\l,\w_n)\;\;\;\;\;\;\;\;\;\;\Rh_n(\l)\equiv\:r(\l,\wh_n) $$
commute with each other:
$$             \U(\l)\U(\mu)=\U(\mu)\U(\l).                            $$

Indeed, the commutation relations of $\w$'s translate
into $N$ copies of the Yang-Baxter equation
$$  \R_{n-1}(\l)\R_n(\l\mu)\R_{n-1}(\mu)
                                 =\R_n(\mu)\R_{n-1}(\l\mu)\R_n(\l)     $$
$$  \R_{n+N}(\l)=\R_n(\l)                                              $$
that is believed to ensure the commutativity of ordered products of those
$R$-matrices. It is not however evident how to make this idea work
in the periodic case where a naive ordered
product, say, $\R_1 \R_2\ldots \R_N$, makes little
sense. The $\p(\star)$ `product' has better
chance to deliver, since, as we know from the preceding section, it does
not at least
depend on the starting point:
$$ \ldots=\p\left( \Rh_0\star \Rh_1\star \ldots \star\Rh_{N-1}\right)
   =\p\left( \Rh_1\star \Rh_2\star \ldots\star\Rh_N\right)=\ldots\;.   $$
And once we get the ordered product right the commutativity check
becomes a matter of familiar $R$-matrix machinery complemented by the
rules of  $\p(\star)$ `multiplication' (and also by the
definition $\hat{\R_n^{-1}}(\l)\equiv\Rh_n(\l^{-1})/\varrho(\l)$ ensuring
that $\p\!\left( \hat{\R_n^{-1}}(\l)\Rh_n(\l)\right)=1$):
$$\begin{array}{rl}
  \U(\l)\U(\mu)\!\!
  &=
   \p\left( \Rh_1(\l)\star \Rh_2(\l)\star \ldots \star\Rh_N(\l)\right)\;
   \p\left( \Rh_0(\mu)\star \Rh_1(\mu)
   \star \ldots \star\Rh_{N-1}(\mu)\right)\\
  &\!\!=
   \p\left(\Rh_1(\l)\star\Rh_0(\mu)\star\Rh_2(\l)\star\Rh_1(\mu)
   \star\ldots
   \star\Rh_N(\l)\star\Rh_{N-1}(\mu)\right)\\
  &\!\!\!\!=
   \p\left( \hat{\R_0^{-1}}(\lm)\star
   \Rh_0(\lm)\star\Rh_1(\l)\star\Rh_0(\mu)\star\Rh_2(\l)\star\Rh_1(\mu)
   \star\ldots\star\Rh_N(\l)\star\Rh_{N-1}(\mu)\right)\\
  &\!\!\!\!\!\!=
   \p\left( \underline{\Rh_0(\lm)\star\Rh_1(\l)\star\Rh_0(\mu)}
   \star\Rh_2(\l)\star\Rh_1(\mu)\star
   \ldots\star\Rh_N(\l)\star\Rh_{N-1}(\mu)
   \star\hat{\R_N^{-1}}(\lm)\right)\\
  &\!\!\!\!\!\!\!\!=
   \p\left( \Rh_1(\mu)\star\Rh_0(\l)\star
   \underline{\Rh_1(\lm)\star\Rh_2(\l)\star\Rh_1(\mu)}\star
   \ldots\star\Rh_N(\l)\star\Rh_{N-1}(\mu)
   \star\hat{\R_N^{-1}}(\lm)\right)\\
  &\!\!\!\!\!\!\!\!\!\!=
   \p\left( \Rh_1(\mu)\star\Rh_0(\l)\star\Rh_2(\mu)\star\Rh_1(\l)\star
   \underline{\Rh_2(\lm)\star}
   \ldots\star\Rh_N(\l)\star\Rh_{N-1}(\mu)
   \star\hat{\R_N^{-1}}(\lm)\right)
                                                            \end{array}$$
$$                           \vdots                                    $$
$$\begin{array}{rl}
  \;\;\;\;\;
  &=
   \p\left( \Rh_1(\mu)\star\Rh_0(\l)\star\Rh_2(\mu)
   \star\Rh_1(\l)\star\ldots
   \star\underline{\Rh_{N-1}(\lm)\star\Rh_N(\l)\star\Rh_{N-1}(\mu)}\star
   \hat{\R_N^{-1}}(\lm)\right)\\
  &\!\!=
   \p\left( \Rh_1(\mu)\star\Rh_0(\l)
   \star\Rh_2(\mu)\star\Rh_1(\l)\star\ldots
   \star\Rh_N(\mu)\star\Rh_{N-1}(\l)\star\Rh_N(\lm)
   \star\hat{\R_N^{-1}}(\lm)\right)\\
  &\!\!\!\!=
   \p\left( \Rh_1(\mu)\star\Rh_0(\l)
   \star\Rh_2(\mu)\star\Rh_1(\l)\star\ldots
   \star\Rh_N(\mu)\star\Rh_{N-1}(\l)\right)\\
  &\!\!\!\!\!\!=
   \p\left( \Rh_1(\mu)\star \Rh_2(\mu)
   \star\ldots \star\Rh_N(\mu)\right)\;
   \p\left( \Rh_0(\l)\star \Rh_1(\l)
   \star \ldots \star\Rh_{N-1}(\l)\right)
   =\U(\mu)\U(\l).
                                                            \end{array}$$

We conclude the Section with two remarks. First, to make formulas easier
on the eyes we shall adopt the
notation ${\op\ldots}$ for `cyclic ordered products' like the ones
above, for instance,
$$           \U(\l)=\op\R_n(\l)                                        $$
$$      \U(\l)\U(\mu)=\op\R_n(\l)\R_{n-1}(\mu)              .          $$
Second, for a reason that will become apparent in Section 6 the
family $\U(\l)$ should be properly `normalized'
$$           \Q(\l)=\frac{\U(\l)}{\U(0)}                               $$
that, of course, does not spoil its commutativity
$$            \Q(\l)\Q(\mu)=\Q(\mu)\Q(\l)                              $$
and polynomiality in $\l$.

\section{Baxter equation}

In the Quantum Inverse Scattering Method language the
family $\Q(\l)$ would
be called the fundamental transfer-matrix [TTF] as opposed to the usual
nonfundamental one which on this occasion has the form [G, V]
$$                   \t(\l)=\tr\op(\W_n L(\l))                         $$
where the matrices $\W_n$ and $L(\l)$ are
$$   \W_n=\bma \w_n^{\half}&\\&\w_n^{-\half} \ema\;\;
                   {\rm with} \;\; \w_n^{\half}\equiv\w_n^{-\ell}      $$
$$                    L(\l)=\bma \l&1\\1&\l  \ema ,                    $$
tr denotes the matrix trace and $\star$ (hidden
in the product symbol) combines what it used to be with
the standard matrix multiplication. In the decyphered form it reads
$$ \t(\l)=\l^{\frac{N}{2}}
          \sum_{k_{1},k_{2},\ldots,k_{N}=\pm\half}
          \q^{2k_{1} k_{N}}\;\l^{2(k_{1}k_{2}+k_{2}k_{3}+
          \ldots+k_{N-1}k_{N}+k_{N}k_{1})}
   \w_{1}^{k_{1}}\w_{2}^{k_{2}}\ldots \w_{N}^{k_{N}}                   $$
with
$$                \q^{\half}\equiv \q^{-\ell}.                         $$
In other words, we have another polynomial in $\l$, this time of
degree $N$\footnote
{To be precise, only odd degrees are present.}
, which, as we know from past experience,
\begin{itemize}
  \item[(i)] commutes with itself
$$             \t(\l)\t(\mu)=\t(\mu)\t(\l),                            $$
  \item[(ii)] commutes with $\Q(\l)$
$$             \t(\l)\Q(\mu)=\Q(\mu)\t(\l)                             $$
\end{itemize}
and, as we are going to see,
\begin{itemize}
  \item[(iii)] satisfies together with $\Q(\l)$ the Baxter equation [Bax]
$$ \q^{N\ell^2}\Q(\l)\t(\l)
                    =\alpha^N (\l)\Q(\q^{-1}\l)+\delta^N(\l)\Q(\q\l)   $$
with
$$                        \alpha(\l)=\l-1                              $$
$$                        \delta(\l)=\q\l+1 .                          $$
\end{itemize}
The first two items of this list are actually superseded by the much
stronger third one the proof of which may go as follows\footnote
{The first proof of (iii) for the model in question was
obtained by R.~Kashaev [K] while the very idea that the fundamental
transfer-matrix $\Q(\l)$ can serve as a Baxter's $Q$-operator probably
belongs to E.~Sklyanin [S]. See also [PG] addressing similar matters.}:
\begin{quote}
$$  \U(\l)\t(\l)=\tr\op(\R_n(\l)\W_{n-1}L(\l))=\tr\op \C_n(\l)         $$
with
$$    \C_n(\l)=B^{-1}\W_{n-1}^{-1}\R_n(\l)\W_{n-1}L(\l)\W_n B,         $$
$$    B=\bma 1&1\\0&1  \ema  .  $$
Using the obvious relation
$$  \w_{n-1}^{\half}\; r(\l,\w_n) \w_{n-1}^{-\half}=\:r(\l,\q \w_n)    $$
one gets
$$  \C_n(\l)=C(\l,\w_n)                                                $$
with
\end{quote}
$$  C(\l,z)
 =\bma \;\;\;z^{\half}(\l \:r(\l,\q^{-1}z)-\:r(\l,\q z))\;\;\;&
         (\l z^{\half}+z^{-\half})\:r(\l,\q^{-1}z)-
         (z^{\half}+\l z^{-\half})\:r(\l,\q z)\\&\\
         z^{\half}\:r(\l,\q z)&
         (z^{\half}+\l z^{-\half})\:r(\l,\q z)   \ema  .               $$
\begin{quote}
The first property of the function $r(\l,z)$ (see Section 2) says
that the  upper off-diagonal element of this matrix
vanishes (provided  $z^{2\ell+1}=1$ and $z^{\half}\equiv z^{-\ell}$).
This yields immediately
$$            \U(\l)\t(\l)=\op a(\l,\w_n)+\op d(\l,\w_n)               $$
where $a$ and $d$ denote the diagonal elements of the matrix $C$:
$$         C(\l,z)=\bma a(\l,z)&0\\ \ast &d(\l,z) \ema .               $$
It remains to verify that
$$  a(\l,z)=\q^{-\ell^2}(\l-1)\:r(\q^{-1}\l,z)                         $$
$$  d(\l,z)=\q^{-\ell^2}(\q\l+1)\:r(\q\l,z) .                          $$
We omit this part of the proof for it is neither difficult nor
instructive.
\end{quote}
To conclude, let us recall what use might be made of the Baxter equation.
The function $\Q(\l)$ is a polynomial in $\l$ with coefficients
coming from the subalgebra of conservation laws. This allows, probably at
the expense of considering a proper completion of the algebra of
observables, the introduction of commuting `roots' of~$\Q(\l)$
$$  \Q(\l)=\prod_k \left(1-\frac{\l}{\r_k}\right)                      $$
$$                 \r_j\r_k=\r_k\r_j      .                            $$
The Baxter equation says, in particular, that
$$  \left.\alpha^N (\l)\Q(\q^{-1}\l)+
                   \delta^N (\l)\Q(\q\l)\right|_{\l=\r_k}=0.           $$
This is just the famous system of the Bethe ansatz equations which
look more  familiar in the form
$$ \left(\frac{\alpha(\r_k)}{\delta(\r_k)}\right)^N=
        -\prod_j \frac{\r_j-\q\r_k}{\r_j-\q^{-1}\r_k}\;\;\;.           $$

\section{Equations of motion}

We are going to see that the generators of
the algebra of observables evolve
$$  \bphi(\l\: |\: \tau,n)\equiv\Q^{-\tau}(\l) \bphi_n \Q^{\tau} (\l)  $$
according to the equations (1) which in ultimately accurate form read
$$ \bphi(\l\:|\:\tau,n)\bphi(\l\:|\:\tau,n-1)
          +\l\bphi(\l\:|\:\tau,n-1)\bphi(\l\:|\:\tau-1,n-1)\;\;\;\;\;  $$
$$\;\;\;\;\; =\l\bphi(\l\:|\:\tau,n)\bphi(\l\:|\:\tau-1,n)
                     +\bphi(\l\:|\:\tau-1,n)\bphi(\l\:|\:\tau-1,n-1) . $$
We shall cover the distance in four short steps.
\begin{itemize}
   \item[(i)] Commutation relations
between $\bphi$'s and $\w$'s have the form
$$           \bphi_n \w_n =\q^2 \w_n \bphi_n                           $$
$$        \bphi_m \w_n=\w_n \bphi_m    \;\;\; m\neq n\mN         .     $$
This prompts us to read the first property of the
function $r$ from Section 2 as
$$   \left(\bphi_{n+1}\bphi_n+\l\bphi_n\bphi_{n-1}\right)\R_n(\l)
   =\R_n(\l) \left(\l\bphi_{n+1}\bphi_n+\bphi_n\bphi_{n-1}\right).     $$
   \item[(ii)]Viewing this equation as a fragment of the `big' one
$$  \sum_{j,k=-\ell}^{\ell} \q^{2jk}\;(\rho_j \w_{m+1}^j)\;\R_{m+2}
      \ldots\R_{n-1}
      \left(\bphi_{n+1}\bphi_n+\l\bphi_n\bphi_{n-1}\right)\R_n \R_{n+1}
                              \ldots \R_{m+N-1} \;(\rho_k \w_{m+N}^k)  $$
$$ =\sum_{j,k=-\ell}^{\ell} \q^{2jk}\;(\rho_j \w_{m+1}^j)\;\R_{m+2}
        \ldots\R_{n-1}
    \R_n  \left(\l\bphi_{n+1}\bphi_n+\bphi_n\bphi_{n-1}\right)\R_{n+1}
                              \ldots \R_{m+N-1} \;(\rho_k \w_{m+N}^k)  $$
and pulling $\bphi$'s to the outside(s) we get
$$   \bphi_{n+1}\bphi_n \U(\l)+\l\bphi_n \U(\l)\bphi_{n-1}
           =\l\bphi_{n+1} \U(\l)\bphi_n+\U(\l)\bphi_n\bphi_{n-1} .     $$
   \item[(iii)] At the point $\l=0$ it turns into
$$        \bphi_{n+1}\bphi_n \U(0)=\U(0)\bphi_n\bphi_{n-1}             $$
suggesting more subtle
$$                    \bphi_n \U(0)=\U(0)\bphi_{n-1}  .                $$
As a matter of fact, the latter does hold. We omit the proof for it is
too case-specific.
   \item[(iv)]  (ii) and (iii) combined yield
$$   \bphi_n\bphi_{n-1} \Q(\l)+\l\bphi_{n-1} \Q(\l)\bphi_{n-1}
                =\l\bphi_n \Q(\l)\bphi_n+\Q(\l)\bphi_n\bphi_{n-1}      $$
which is nothing but (1) with cut away common
factors $\Q^{-\tau}(\l) \ldots\Q^{\tau-1} (\l)$.
\end{itemize}
So, we have met the last objective of the Letter.
We conclude it with two remarks.

$\bullet$ A consistent approach to the subject should probably
distinguish between observables and their automorphisms rather
than mix them up as we did in this Letter. It would be only natural
to deal not with the $R$-matrices but directly with
automorphisms they represent:
$$    \aR_n(\l):\begin{array}{rcl}
 \bphi_n&\mapsto&\frac{1+\l\:\q\w_n}{\l+\q\w_n}\bphi_n\\
 \bphi_m&\mapsto&\bphi_m \;\;\; m\neq n\mN \end{array}
\;\;\;\;\;. $$
Indeed, it would follow straight from the above definition that
$$     \bphi_{n+1}\bphi_n+\l\bphi_n\bphi_{n-1}
         \stackrel{\aR_n(\l)}{\longmapsto}
          \l\bphi_{n+1}\bphi_n+\bphi_n\bphi_{n-1}                      $$
and
$$  \aR_{n-1}(\l)\circ\aR_n(\l\mu)\circ\aR_{n-1}(\mu)
                   =\aR_n(\mu)\circ\aR_{n-1}(\l\mu)\circ\aR_n(\l)  .   $$
These two relations are `weaker' (but just sufficient!) substitutes
for the two cornerstones of the whole scheme, which are item (i) of
this Section and the Yang-Baxter equations of Section 4. Why then care
whether those `$R$-automorphisms' are inner or not? They happen
to be inner in our particular case but even there some other important
automorphisms are outer anyway [FV93]. In other cases one pays a
dear price for `inclusion' of $R$-matrices in the algebra of
observables [BR, F, FV95]. Unfortunately, some difficulties of
the `automorphism' approach made the author to choose for this
Letter the more familiar `inner' route.

$\bullet$ It would be useful to know whether the equations of
motion (1) provide the exhaustive information about the model.
In other words, is $\bphi(\tau,n)$ the only solution
to the Cauchy problem
$$              \bphi(\tau,n)|_{\tau=0}=\bphi_n                        $$
for the system (1)? Not quite, and it is easy to
see why. First, in our solution the `quasimomentum' $\c$ (see
Section 1) does not evolve but the equations (1) do not know
about that. Also, they do not feel whether we
multiply $\R_n(\l)$'s `from left to right' or $\R_n(-\l)$'s in the
opposite
order. As a matter of fact, there are no other sources of nonuniqueness.
So, the local structure of (1) is good and all one needs to achieve
the ultimate uniqueness is a
couple of extra global conditions. This point
will be described in more detail elsewhere.
\bigskip
\begin{quote}
{\bf Acknowledgements.}
I greatly appreciate the hospitality of Department of Theoretical
Physics of Uppsala University.
I would like to thank
O.~Babelon, L.~Faddeev, B.~Feigin, M.~Flato, R.~Kashaev,
A.~Kirillov, J.-M.~Maillet, A.~Reyman,
N.~Reshetikhin, M.~Semenov-Tian-Shansky and especially E.~Sklyanin for
helpful discussions.
\end{quote}
\section*{\normalsize\bf References}

\begin{itemize}
  \item[{[B]}] O. Babelon, Phys. Lett. B238 (1990) 234.
  \item[{[Bax]}] R. Baxter, Exactly Solved Model in Statistical
Mechanics (Academic, New York, 1982).
  \item[{[BR]}] V. Bazhanov and N. Reshetikhin, private communication.
  \item[{[F]}] L. Faddeev, hep-th/9406196.
  \item[{[Fe]}] B. Feigin, private communication.
  \item[{[FV92]}] L. Faddeev and A. Volkov,
Theor. Math. Phys. 92 (1992) 207.
  \item[{[FV93]}] L. Faddeev and A. Volkov, Phys. Lett. B315 (1993) 311.
  \item[{[FV94]}] L. Faddeev and A. Volkov,
Lett. Math. Phys. 32 (1994) 125.
  \item[{[FV95]}] L. Faddeev and A. Volkov, Zap. Nauchn.
Semin. PDMI 224 (1995).
  \item[{[FT]}] L. Faddeev and L. Takhtajan, in:
Lecture Notes in Physics 246 (Springer, Berlin, 1986) 66.
  \item[{[FZ]}] V. Fateev and A. Zamolodchikov,
Phys. Lett. A92 (1982) 37.
  \item[{[G]}] J.-L. Gervais, Phys. Lett. B160 (1985) 279.
  \item[{[H]}] R. Hirota, J. Phys. Soc. Japan 43 (1977) 2079.
  \item[{[K]}] R. Kashaev, private communication.
  \item[{[NC]}] F. Nijhoff and H. Capel, Acta Appl. Math. 39 (1995) 133.
  \item[{[PG]}] V. Pasquier and M. Gaudin, J. Phys. A25 (1992) 5243.
  \item[{[S]}] E. Sklyanin, private communication.
  \item[{[TTF]}] V. Tarasov, L. Takhtajan and L. Faddeev, Theor. Math.
Phys. 57 (1983) 163.
  \item[{[V]}] A. Volkov, Phys. Lett A167 (1992) 345.
\end{itemize}

\end{document}